\documentclass[a4paper,twoside]{article}

\usepackage{epsfig}
\usepackage{subcaption}
\usepackage{calc}
\usepackage{amssymb}
\usepackage{amstext}
\usepackage{amsmath}
\usepackage{amsthm}
\usepackage{multicol}
\usepackage{pslatex}
\usepackage{apalike}
\usepackage{booktabs}
\usepackage[bottom]{footmisc}
\usepackage[english]{babel}

\usepackage[letterpaper,top=2cm,bottom=2cm,left=3cm,right=3cm,marginparwidth=1.75cm]{geometry}

\usepackage{amsmath}
\usepackage{graphicx}
\usepackage[colorlinks=true, allcolors=blue]{hyperref}

\usepackage{SCITEPRESS}

\begin{document}

\title{Student's attraction for a carrier path related to Databases and SQL. Usability vs Efficiency in students' perception. Case Study.}

\author{\authorname{Manuela-Andreea Petrescu\sup{1}\orcidAuthor{0000-0002-9537-1466}, 
Emilia-Loredana Pop\sup{1}\orcidAuthor{0000-0002-4737-4080}}
\affiliation{\sup{1}Department of Computer Science, Babes Bolyai University, Cluj-Napoca, Romania} 
\email{manuela.petrescu@ubbcluj.ro, emilia.pop@ubbcluj.ro}
}


\abstract{ 
This study explores and analyses the expectations of second-year students enrolled in different lines of study related to Database course, as their interest in having a carrier path in a database related domain and how it reflects the job demands from the market. The participants in the study provided two sets of answers, anonymously collected (in the begging and in the middle of the course), thus allowing us to track how their interests changed as long as they found out more about the subject. We asked for their experience and initial knowledge, we found out that they are aware of the SQL and databases' usability and importance, but they appreciated the database knowledge will be used occasionally. Even if it was not the original scope of the paper, we also found out that men are more interested in learning in-depth (acquiring security, performance, complexity database related information) than women do. In terms of the participants set, there were 87 answers from 191 enrolled students that were analyzed and interpreted using thematic analysis. }

\onecolumn \maketitle \normalsize \setcounter{footnote}{0} \vfill


\section{Introduction}
Nowadays, databases mean power, usage, freedom, and plenty of possibilities in the real life. Database (DB) plays an important role in today's society, in all aspects of modern life, being utilized by companies, public and private sectors of activity, research fields of activity, personal life, and students' career paths. Databases play a major role in student management at universities all over the world. 

The database impact is huge not only in the academic world, but also in the medical world (databases that store images for medical purposes as radiographers), in the car sector (the databases are also used for image recognition), in the military sector (there are databases with maps), and so on, even if the classical databases evolved (NoSQL databases for example). Because of this demand and high database impact, there is a need for specialized personnel, so SQL and database know-how become a must for the students in computer sciences. However, the domain attractiveness (Computer Science and/or Mathematics) for secondary schools students depends on a set of factors: the student's socioeconomic status, motivation, performance, self-efficacy, task value beliefs, engagement in Computer Science \cite{Kahraman22,Spieler20}. Also, female students should be empowered through direct encouragement, female-only initiatives, and mentoring programs according to the same authors. 




We wanted to analyze the student's perception and their interest in following a database-related carrier, and if the student's interest decreases or increases once they know more about this field of work. Similar studies involving a variety of disciplines were performed by \cite{Kinash17} in Australia, and a deeper understanding and relevant progression paths were given by \cite{Gaebel12} and \cite{Allum14} (for PhD career pathways that can involve databases).

We performed our analysis related to the Database course involving students from different specializations and lines of study.

The scope of the paper is to find out what are the students' expectancies about working with Databases and SQL related domains and how these aspects correlate with the demand in the labor market. We wanted to find out how interested are they at the beginning of the course (first survey), correlating the interest level with DB/SQL knowledge (as some of them have previous experience, others did not work with an SQL statement) and if the interest changed during the course (we performed a mid-term quiz - second survey).  We also analyzed if they prefer a career path that involves deeper usage of SQL (or NoSQL statements), database security, or efficiency, or they just consider these as being of seldom usage in their career path or future job. To find out, we asked them open questions: \textit{R1: Do you have DB-related knowledge?, R2: What are your expectations related to DB course? What do you want to learn?} and \textit{R3: How do you plan to use the learned information? Are you interested to work in DB-related fields?} 

Even if it was not the initial purpose of our paper, we investigate how the students related themselves to usability vs efficacy in using databases in men compared to women. 

The paper is structured as follows: it starts with an introduction and a \textit{Literature Review} section where we analyzed other papers that discussed the same topic. In the \textit{Methodology} section, we talked about the methods used in this survey, starting with the participant's set, curricula, and asked questions. Survey questions and data analysis was performed in the section \textit{Data Collection and Analysis}, where we analyzed the received responses, performed a comparison between men's and women's goals and we have analyzed the job taxonomies based on public ads versus students' interest to work in such a position. We have analyzed the possible \textit{Threats to Validity} and the actions performed to minimize and mitigate them. The \textit{Conclusion and Future Work} section ends and summarises our work and the obtained results, and mentions future approaches and work.


\section{Literature Review}
\label{sec:literature_review}
The databases topic proved to be important and useful in the last decade, as the storage capacity become cheaper, the databases increased their size, storing more and more information, and bringing different perspectives and alternatives to people's carriers. 

In parallel with computer science (and database related knowledge) development, women's activity in this area increased, due to the variety of possibilities to which computer science and database knowledge can be applied. Different job positions appeared to satisfy the demand from the computer science domain, and as more jobs were generated, women found it easier to work in a field that was previously dominated by men \cite{Weston19,Spieler20,Berkeley21}. The variety of available positions jobs in database-related field development include Database Administrator, Data Scientist/ Data Analyst, Data Operator, Database Migration Operator, Data Migration Manager, Relational/ Non-relational/ SQL/ NoSQL  Database Developer ( Azure Lead Developer Backend), Business Analyst/ Business Intelligence Developer, Database Manager, Software Developer/ Database Engineer, Database Architect, and many others (for example \cite{Jaiswal22}). 

In the idea of choosing the best career path for students, the article \cite{Verma17} presents an useful and comprehensive student-centric recommendation system. The article is based on the research analytics framework and uses the three-dimensional model for measurements integrated with the relative weighted set generated using the Analysis Hierarchical Process decision system. Other guidance in choosing the career path is given by \cite{Hasan20} and presents also psychological, sociological, and developmental perspectives. It also reveals the needs of the individual for a sustainable future, taking into consideration external factors, internal improvements, path-related dynamism, and practical aspects. Aspects related to the idea of choosing the desired career path with a high degree of efficiency and a comparison between genders are presented in \cite{Mann20}. Digital carrier aspiration seems to be influenced by gender, according to \cite{Wong18}, who performed 32 semi-structured interviews with digitally skilled teenagers, aged from 13 to 19 years old, and analyzed and identified their digital interests. 
Creativity was identified as being a career pathway very important and especially the girls proved to tend to become a consumer rather than a creator in technology. 
\cite{Mckenzie22} investigated the career aspirations of undergraduate IT students from an Australian university, by completing an online self-assessment of study and career confidence related to the discipline and also a survey about short-term and longer-term career aspirations and prior experience. The intrinsic interest and enjoyment of IT proved to be the best motivation to study IT and also brought insights for career aspirations, without realizing the time need it to such a job position. Combining the presentation of an alumni database architecture that could extract, transform and load data from alumni LinkedIn profiles of the students who graduated with Bachelor of Science degrees and Master of Science degrees with their career trajectory, the conclusions showed, for them, the choice of different career paths, like engineers and managers on one side and managers and analysts on the other side \cite{Li16}. In the literature review, there are studies with important elements and arguments related to databases and also surveys performed to obtain relevant information, from students, professors, and different members \cite{Dehghani18,Uzun20}. 

One of the most important aspects related to databases is connected to the Structured Query Language (SQL), which provides relevant information for a potential user. In Computer Science and Software Engineering, SQL skills are mandatory. Besides the knowledge of the SQL language, pedagogical skills are needed in education. \cite{Taipalus20} presents an overview of the educational SQL research topics, research types, publication fora, and also some SQL teaching practices. The idea for educational SQL research should be to include replication studies, studies on advanced SQL concepts, and studies on aspects that are not related to data retrieval. This study includes also teaching practices in SQL education and a systematic map of educational SQL research and future research agenda. 

Databases include big amounts of data stored in complex and different manners with the help of various methods and tools. Databases can include SQL, NoSQL, Big Data, Business Intelligence, Data Science, or Data Analytics. 
\cite{Fotache15} presents some of the coordinates used in processing data and implications for the academic curricula, and also provides arguments for the positions of Data Analyst and Business Intelligence to acquire a corresponding level of SQL and  Data Warehouses knowledge. 

Nowadays, there are plenty of tools and online trainings, documents (see, for example, \cite{IBM10}, \cite{Halvorsen17}), articles, and video's that present in a detailed manner, aspects related to databases, such that each of us can learn the SQL / NoSQL technique. 


\section{Methodology}


The survey research method was used for the study, according to ACM Sigsoft Empirical Standards for Software Engineering Research \cite{ACM}. The questionnaire included two one-choice questions and two open questions. The idea was to obtain multiple and relevant information. 


In total, there have been given two questionnaires to the students, one at the beginning of the semester and another one in the middle of the semester. The idea was to measure the differences that appeared in students' perceptions related to a carrier tightly connected to the course topic, at the beginning, when the subject was relatively unknown to them, and in the middle of the semester, when the subject began to be known. 
This aspect can reveal the involvement of students in database course. 
We repeated some of the questions to check how the student's interests changed after they acquired Database knowledge. The students that took part in the study were enrolled in the Database course in different lines of study, but the learning topics (course syllabus) were identical. The topics for the courses are the ones given in Table \ref{tab:curricula}.
\begin{table}
  \caption{\textit{Databases} curriculum}
  \label{tab:curricula}
  {\small{
  \begin{tabular}{p{7.0cm}}
\toprule
Curriculum\\ 
\midrule
Lecture 1. Introduction to Databases. Fundamental Concepts. \\
Lecture 2. The Relational Model. \\
Lecture 3. SQL Queries.\\
Lecture 4. Functional Dependencies. \\
Lecture 5. Normal Forms. \\
Lecture 6. Relational Algebra.\\
Lecture 7. The Physical Structure of Databases. \\ 
Lecture 8. Indexes. \\ 
\toprule
\end{tabular}
}}
\end{table}


The questions from the surveys have been anonymous and optional, a student could answer one question, to all of them or none. Because of this aspect, we have fewer participants in the study (students that provided feedback) compared to the total number of students: 87 vs 191 enrolled students. The closed questions from the questionnaire relieved the specialization and the gender of the students and were used to determine the participant groups while performing the data analysis.

\subsection{Participants}

We asked a number of 191 enrolled students to participate in the survey, from which a number of 87 participated and provided answers.  

We considered that the number of students that answered is representative in terms of percentage 46\% and in terms of received number of responses. The received number of responses is comparative with the number of received responses from other published studies in the computer science domain \cite{icsoft22,enase21}.  
The students asked to answer the survey were from the following specialization and lines of study:
\begin{itemize}
\item Computer Science - Romanian line, 84 students from which 61 men and 23 women,
\item Computer Science - English line, 52 students from which 41 men and 11 women, 
\item Mathematics and Computer Science - English line, 55 students from which 36 men and 19 women. 
\end{itemize}

191 students were enrolled in the course, 53 girls and 138 boys, the percentage of the enrolled girls is relatively similar for all lines of study; the women vs men percentage of enrolled students is 28\%. We wanted to make sure that the survey is representative for all the segments of the population, so we asked for gender and we got 42\% of the answers stating there are girls. Due to these aspects, we concluded that the girls were well represented in the responses received from the target population. Women seem to be more likely to participate in the survey.  

\subsection{Data Collection and Analysis}
\label{sec:data_collection}
The responses used in this study have been collected anonymously, using Google forms, thus allowing us to see in real-time the provided answers and the time when they were submitted. Most of the responses were submitted on the same day (45,54\% of the total number). The same students were required to participate in both surveys (after the first course and in the middle of the semester), and they were informed about the purpose of the surveys and how the provided information will be used. The surveys remained open for 10 days allowing everyone who was interested to participate to submit their point of view. We opted for closed questions to be able to establish data sets more precisely and for open questions for the topics where we wanted to have more data and a profound understanding. For collecting data we used quantitative methods: in fact, specific questionnaire surveys as they were specified in the empirical community standards \cite{ACM}; questionnaire surveys were used in other computer science related studies \cite{tichy95,redmond13,easeai22,icsoft22}. For text interpretation we used thematic analysis \cite{Braun19} (previously used in computer science studies in \cite{Kiger20,enase21}), and performed the following steps: 
\begin{itemize}
\item Obtaining the answers.
\item If needed, the answers were restructured and allocated to other study questions, as parts of the answers were sometimes better fitted as an answer to other questions.
\item Determined the specific keywords.
\item Selected keywords were grouped into classes by one author.
\item The other author verified the classification, and did some observations; the observations were analyzed by both authors and small changes were done. 
\end{itemize}
Sometimes the students mentioned their SQL knowledge as an answer to the course expectations, or there were other cases when the answer for a question was detailing aspects of another question. Due to these facts, we had to remap some answers (or parts of the answers). Some answers contained one keyword, and others contained more than one; so we decided to analyze the percentage of answers where a specific keyword appears, as a consequence, we will work with percentages.

For the first quiz, we gathered 52 answers from students and 37 answers for the second quiz. The questions asked in both surveys - common question are QC1,QC2, QC3 and QC4, in the first quiz we also had Q1 and Q2:
\begin{itemize}
\item QC1: Line of study (choice of \textit{Computer Science / Mathematics and Computer Science})
\item QC2: Gender (Choice of \textit{Male / Female / I don't want to answer}).
\item QC3: How do you plan to use the learned information?
\item QC4: Are you interested to work in DB-related fields?
\item Q1: Do you have DB-related knowledge? 
\item Q2: What are your expectations related to the Database course?
\end{itemize}

\subsection{Q1: Do you have DB-related knowledge?}
This course was an introductory lecture to databases and SQL statements, so we expect to have enrolled students that did not work previously with SQL or with databases. We wanted to find out what is their knowledge level, and consequently, we had students that did not have any SQL experience or knowledge: \textit{"My database knowledge is practically nonexistent", "I have NoSQL or databases knowledge"} and we had students that mentioned \textit{"I had an internship as a software engineer where I worked with SQL", "I currently use some of the principles at work", "I used SQL queries working on a SpringBoot internship program"}. However, none of the students recommended themselves to have \textit{good} or \textit{very good} database knowledge. In terms of self-evaluation, 19.23\% of them mentioned they don't know SQL or database related information and a similar percentage, 21.15\% mentioned that have little knowledge. We could conclude that the knowledge level of database related topics was relatively low among the students that enrolled in the course, aspects that can be correlated with the fact that around 84\% of them mentioned that they want to learn databases basics.

\subsection{Q2: What are your expectations related to the Database course?}
\begin{figure}[htbp!]
    \includegraphics[width=0.5\textwidth]{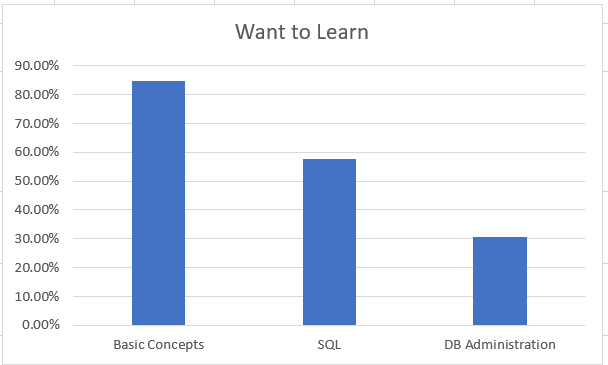}
    \caption{Student's Interests}
    \label{fig:wantToLearn}
\end{figure}

We found three keywords related to \textit{database basic concepts} - 84\%, \textit{database administration} - 31\%, and \textit{SQL learning} - 58\% as depicted in Figure \ref{fig:wantToLearn}. A large part of the students just wanted to find out and understand how databases are working, to be able to use them without knowing too much about complexity, efficiency, or security.

Most of the students were aware of the databases' role and importance and expressed their interest in learning the basics: \textit{"Understanding the databases and their role in applications", "To know how databases are functioning", "To learn to create a database having a good table structure, the SQL statements are not important as I can find them on internet. We should know only that they exist, how are called and used for, and then we could search for them on the internet"}.
Even if we had a couple of answers stating their expectations are pretty high, as they want to be able to operate without problems complex databases, we concluded that on average, the student's expectations for the course were to achieve only basic knowledge and most of them were not interested in topics such as performance, efficiency, security, and so on.

\subsection{Q3: How do you plan to use the learned information? Are you interested to work in DB-related fields?}

In the first quiz, 42,30\% of the students mentioned that they plan to use SQL and DB knowledge only occasionally, as they consider DB knowledge to be useful in developing an application: \textit{"I want to learn DB theory and to apply it in an application", "I want to learn new things that could help me in the future" or "I'll need this information when I will develop my applications/sites"}.

In the middle of the semester, in the second quiz, 27\% of the students considered that the course information is interesting, and useful in every domain appeared in 17\% of the answers. 11\% of the answers mentioned they are more attracted to the DB domain. However, in total, the percentage of students that would not choose to work in a DB-related field was 55,56\%, compared to 25\% in the first quiz. The percentage of students that did not decide decreased from 26.9\% in the first quiz to 11.11\% in the second quiz, so as they learned more about databases, the percentage of students that don't want to work in a DB-related field increased. The percentage of the students that want to work in DB-related fields remained relatively the same, as the students that were undecided made up their minds by not preferring a DB-related carrier path.


There is an interesting aspect related to the Database course versus having a career related to databases, as even if the students find the course interesting, they do not picture themselves having a career in this domain: \textit{"I found this course interesting, but I keep my opinion: I don't think I'll work in the Database domain, but I'm 100\% sure that I'll use them", "I like the idea of databases, but I don't see myself working in this field, not at least directly", "I see it more clearly and the approach is not very complex so I can acquire some confidence in this subject. I don't exclude working in this field of activity but I'm not sure for now."}
We concluded that the number of students wanting to work in a DB-related field was constant compared to the beginning of the course, and the number of students that did not want to work in a DB-related field grew as the number of undecided students shrunk. Also, most of the students are aware that databases knowledge is a must when working as a programmer, but they plan to use it only occasionally.

\subsection{Usability versus efficiency and complexity, expectations in men vs women}
\begin{figure}[htbp!]
    \includegraphics[width=0.48\textwidth]{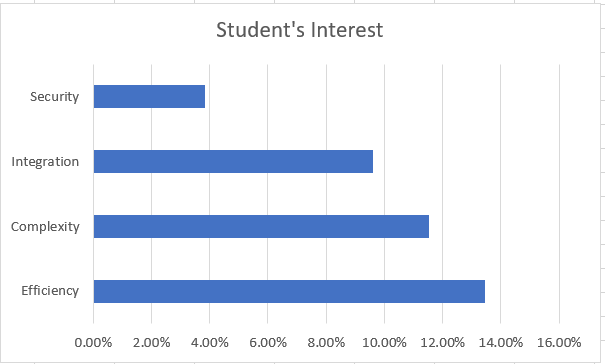}
    \caption{Student's Interest}
    \label{fig:stInterest}
\end{figure}

Even if it was not a study question, when we analyzed the responses to the questions, we noticed that the men's answers show different expectations compared to women's and we performed a comparison between them. Most of the students did not set high goals: 84\% mentioned that they want to know the basics, and only 13,46\% mentioned efficiency as an achievable goal, {\em to retrieve the information from databases in an efficient manner (fast)"}. Complexity appeared in 11,54\% of the total number of answers in the response related to course expectations question: {\em "I want to learn to manage a database in a complex application", "I expect that by the end of the course to be able to administrate and manipulate a database regardless of its complexity"}. Security was not on the list of priorities, only 3,85\% mentioned it: \textit{"how to protect the data and to make a database more efficient"}. Integration with other applications or development environments appeared in 9.72\% of the answers: {\em "to know how a database works, what are the relations between them and how to integrate it in different programming languages", "how to access them from an external programming language(C\#)"}. 
As a note, some answers  mentioned one or more aspects related to either security, complexity, or integration, some students were more aware of these aspects than others, but the overall percentage did not exceed 14\% of the total answers, it can be visualized in Figure \ref{fig:stInterest}.


Related to gender, we wanted to find out how the students manifest their interests and we have seen that men manifested a preoccupation for complexity, security, performance, and integration more than twice compared to women. The following figure (Figure \ref{fig:stInterestGender}) presents the difference between men and women (some students did not specify their gender, and without considering them, the percentage of men vs women preoccupation varies between 200\% and 300\%).

 \begin{figure}[htbp!]
    \includegraphics[width=0.48\textwidth]{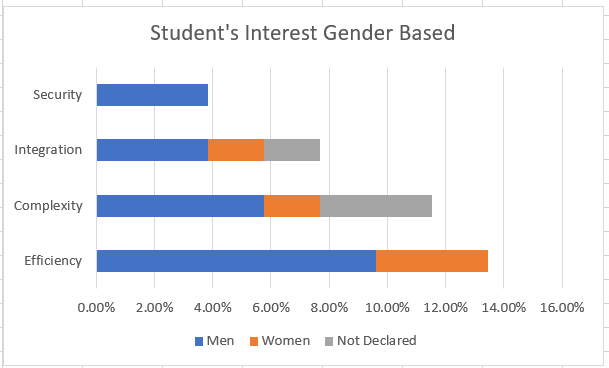}
    \caption{Student's Interest by Gender}
    \label{fig:stInterestGender}
\end{figure}

In this context, other answers also raised our interest, as they mentioned they expect to learn \textit{"without stress"} or \textit{"to know on an average level"}. Based on the data gathered and on their responses, we can conclude that the students are not interested in learning databases information at a high and complex level, preferring a more basic level. Also, men seem to be more interested in learning more advanced features related to security, complexity, or efficiency compared to women.

\subsection{Mentioned Jobs versus Jobs Taxonomy in database-related fields}
The participants in the study were students in the second year of Computer Science, due to this aspect, we presumed that they would mention a larger variety of jobs in their answers. However, it turned out that they were relatively unaware of the jobs that require SQL or DB administration skills, some of them stating that they do not know what is SQL. They just wanted to be able to use databases and SQL instructions when needed, on an occasional base. 
We turned to computer science literature \cite{Bennett22} and technical reports \cite{RePEc22,ESCoE18} that evaluated the jobs and the required skills based on the job ads. In \cite{Bennett22}, SQL skills appeared tightly related to "Programming language or specialized software, Java, SQL, Python", technical reports \cite{RePEc22} mentioned only one job related to SQL and databases: 'SQL Database Administrator' that appeared in 0.49\% of announcements. \cite{ESCoE18} performed a more detailed analysis of the job taxonomy and required skills, SQL skills appear in 8.37\% proportion of unique job adverts, SQLite appeared in 0.17\%, and it was related to application development. Related to BI (Business Intelligence) and Data warehousing, jobs like  java developer, data architect, consultant, analyst, and Oracle developer required database administration skills; the ads were in 1.37\% proportion. There are also Data Engineer related jobs (NoSQL, optimization, MongoDB, or PostgreSQL skills are required), but the percentage is relatively low: 0.72\% compared to software development ads proportion: 5.29\%.

Based on these results, it seems that students managed to focus on the most important part - SQL knowledge in trend with the labor market demand.

\section{Threats to Validity}

We paid attention to mitigate the validity risk, and based on the guidelines defined in \cite{ACM}, we decided to focus and analyze the following topics: target participant set, participants selection, contingency actions for drop-outs, and we also took into consideration research ethics. 

The target participant set was represented by a group of students enrolled in the database course in different specializations (\textit{Computer Science} and \textit{Mathematics and Computer Science}) and in different lines of study (Romanian and English). 
The participant selection was done using groups of study, and the student's grouping was done in alphabetical order. After the groups were randomly selected, there was no other participant selection, so all the students belonging to the study groups were included in our study. Thus there were no threats related to the participant set or participant selection.

As participation in the study was optional, our methods to enlarge the participation were limited; except for explaining the survey purpose and asking students to participate, there was not much we could do. We could not enforce participation, even though it was lower in the second survey compared to the first one (37 versus 52 responses).

As for research ethics, we informed the students about the purpose of the questionnaire survey, and about the fact it is optional (proved by response rates). The survey questions were not mandatory, so a student could answer only one or could answer all the questions. 

We had to address the possibility that our approach to data processing was a subjective one. We tried to mitigate the risk by following the recommendation for data processing and, as authors, by checking each other's work.  

\section{Conclusion and Future Work}
We planned to analyze how second-year students enrolled in Computer Science and Mathematics and Computer Science, perceived the SQL and databases importance and how attracted are they to work in a related domain. Even if a number of 191 students participated in the course, only a part of them effectively participated in the survey by providing answers. We asked for their expectation and interest to work in a database-related field at the beginning of the course and we checked their interest in the middle of the course. We tried to eliminate all the possible threats to validity by having a diverse set of participants (in terms of line of study and gender - all students enrolled in the course could participate if they wanted), an anonymous survey, and analyzing the text responses as recommended by community standards. We found out that they wanted to learn SQL at a medium level. By the middle of the semester, students that were undecided at the beginning of the course, mentioned that they did not want a job related to databases (DB administrator, SQL developer, and so on). Their options for following a carrier path in database-related domains match the job offerings from the market. 
Even if was not the purpose of the paper, we found out that men are more interested in accumulating knowledge and having a more profound understanding of the concepts in aspects related to security, complexity, and efficiency compared to women.
In the future, we plan to find out if the same trend in gender differences persists at the end of the course and to verify these aspects in other courses, and to extrapolate this study to the university degree as a whole.\\

{\bf Funding}
The publication of this article was supported by the 2022 Development Fund of the Babe\c{s}-Bolyai University.

\bibliographystyle{apalike}
{\small
\bibliography{main}}


\end{document}